\documentclass[twocolumn,showpacs,amsmath,amssymb,aps,
prd,nofootinbib]{revtex4}

\usepackage{graphicx}

\begin{document}

\title{Nonlinear evolution of r-modes: the role of differential
rotation}

\author{Paulo M. S\'a}

\email{pmsa@ualg.pt}

\author{Brigitte Tom\'e}

\email{btome@ualg.pt}

\affiliation{Departamento de F\'{\i}sica and Centro
Multidisciplinar de Astrof\'{\i}sica -- CENTRA, \\ F.C.T.,
Universidade do Algarve, Campus de Gambelas, 8005-139 Faro,
Portugal}

\date{November 15, 2004}

\begin{abstract}
Recent work has shown that differential rotation, producing large
scale drifts of fluid elements along stellar latitudes, is an
unavoidable feature of \textit{r}-modes in the nonlinear theory.
We investigate the role of this differential rotation in the
evolution of the $l=2$ \textit{r}-mode instability of a newly
born, hot, rapidly rotating neutron star. It is shown that the
amplitude of the \textit{r}-mode saturates a few hundred seconds
after the mode instability sets in. The saturation amplitude
depends on the amount of differential rotation at the time the
instability becomes active and can take values much smaller than
unity. It is also shown that, independently of the saturation
amplitude of the mode, the star spins down to rotation rates that
are comparable to the inferred initial rotation rates of the
fastest pulsars associated with supernova remnants. Finally, it is
shown that, when the drift of fluid elements at the time the
instability sets in is significant, most of the initial angular
momentum of the star is transferred to the \textit{r}-mode and,
consequently, almost none is carried away by gravitational
radiation.
\end{abstract}


\maketitle

\section{\label{int}Introduction}

The interest in \textit{r}-modes, first studied more than twenty
years ago \cite{pp}, has increased dramatically when it was
discovered \cite{and}, and afterwards confirmed more generally
\cite{fm}, that \textit{r}-modes are driven unstable by
gravitational radiation reaction in perfect-fluid stars with
arbitrary small rotation. Soon afterwards, it was shown
\cite{lom,aks} that in a newly born, hot, rapidly rotating neutron
star, bulk and shear viscosity do not suppress the \textit{r}-mode
instability for a wide range of relevant temperatures and angular
velocities of the star. As a result, the neutron star could spin
down to just a small fraction of its initial angular velocity,
thus providing a possible explanation for the relatively small
spin rates of young pulsars in supernova remnants \cite{lom,aks}.
The gravitational waves emitted by the young neutron star during
this spin-down phase could be detected by enhanced versions of
laser interferometer detectors \cite{olcsva}. It was also
suggested that in accreting neutron stars in low-mass x-ray
binaries the gravitational-wave emission due to the
\textit{r}-mode instability could balance the spin-up torque due
to accretion, thus limiting the maximum angular velocity of these
stars to values consistent with observations \cite{bil,akst}.

Of fundamental importance in judging the astrophysical relevance
of the \textit{r}-mode instability is the determination of the
saturation amplitude of the mode. In recent years, several authors
have addressed this issue, both analytically and numerically,
taking into account different nonlinear effects
\cite{rls,rlms1,rlms2,sf,ltv,ltv2,glssf,afmstw}.

An approximate analytical expression for differential rotation
induced by \textit{r}-modes was first derived using the linearized
fluid equations by expanding the velocity of a fluid element
located at a certain point in powers of the mode's amplitude,
averaging over a gyration, and retaining only the lowest-order
nonvanishing term \cite{rls,rlms1}. This differential rotation can
then interact with the magnetic field of a newly born, hot,
rapidly rotating neutron star, limiting the growth of the
\textit{r}-mode instability or, for strong magnetic fields, even
preventing it from developing \cite{rls,rlms2}.

The first numerical study of nonlinear \textit{r}-modes was
performed for a rapidly-rotating relativistic star without a
gravitational radiation force \cite{sf}. These studies show there
is no suppression of the mode, even when its initial amplitude is
of order unity, and also confirm the existence of differential
rotation induced by r-modes. In a subsequent numerical study of
the nonlinear evolution of \textit{r}-modes of a rapidly-rotating
Newtonian star with a gravitational radiation force, it was shown
that the amplitude of the mode grows to order unity before strong
shocks near the stellar surface rapidly damp the mode
\cite{ltv,ltv2}. However, a more recent numerical simulation has
found no evidence that the decay of the mode's amplitude is due to
such shocks near the surface of the star; instead, the
catastrophic decay of a \textit{r}-mode's amplitude of order unity
is due to a leaking of energy into other fluid modes, leading to a
differentially rotating configuration \cite{glssf}.

Recently, the nonlinear coupling between stellar inertial modes
has been analyzed, with the conclusion that \textit{r}-modes may
saturate at much lower values than previous investigations had
revealed \cite{afmstw}. Despite this fact, it was found that the
\textit{r}-mode instability could still explain the spin
clustering at the fast end of the spin distribution of neutron
stars in low-mass x-ray binaries and that gravitational waves from
newly born, hot, rapidly rotating neutron stars, as well as from
old neutron stars in low-mass x-ray binaries, could be detected by
enhanced versions of laser interferometer gravitational wave
detectors \cite{afmstw}.

In this paper, we are concerned with the role of differential
rotation in the evolution of the \textit{r}-mode instability.

Recently, a nonlinear extension of the linear \textit{r}-mode
perturbation was found within the nonlinear theory up to second
order in the mode's amplitude $\alpha$ in the case of a Newtonian,
barotropic, perfect-fluid star rotating with constant angular
velocity $\Omega$ \cite{sa}. This nonlinear extension describes
differential rotation of pure kinematic nature that produces large
scale drifts along stellar latitudes. This solution contains two
separate pieces, one induced by first-order quantities and another
determined by the choice of initial data. These two pieces cannot
cancel each other, since one is stratified on cylinders and the
other not. Thus, differential rotation is an unavoidable kinematic
feature of \textit{r}-modes.

The differential rotation found in Ref.~\cite{sa} does contribute
to the second-order physical angular momentum of the
\textit{r}-mode, implying that the phenomenological model proposed
in Ref.~\cite{olcsva} to study the evolution of the
\textit{r}-mode instability in newly born, hot, rapidly rotating
neutron stars has to be modified so that the physical angular
momentum of the perturbation includes, not only the canonical
angular momentum, but also the contribution from second-order
differential rotation. In this paper, we study such a modified
model, arriving at the conclusion that differential rotation plays
an important role in the evolution of the \textit{r}-mode
instability.

In section \ref{dif-rot} we derive an analytical expression for
the physical angular momentum of the r-mode perturbation, using
the second-order solution of Ref.~\cite{sa}. In section
\ref{model}, the model for the evolution of the \textit{r}-mode
instability is presented and a system of nonlinear, coupled,
differential equations for the mode's amplitude $\alpha$ and the
star's angular velocity $\Omega$ is derived. In section
\ref{role}, the analytical solution of the above mentioned system
of differential equations is presented, revealing that
differential rotation does play an important role in the evolution
of the \textit{r}-mode instability. Finally, in section \ref{con}
we present the conclusions.

\section{\label{dif-rot} Physical angular momentum of the
\textit{r}-mode perturbation}

For small-amplitude perturbations of slowly rotating, Newtonian,
barotropic, perfect-fluid stars, the \textit{r}-mode solution to
the linearized Euler and continuity equations is, in spherical
coordinates $(r,\theta,\phi)$, given by:
\begin{subequations}
\label{vf}
\begin{eqnarray}
\hspace{-0.6cm} \delta^{(1)} v^r &=& 0,
    \label{vf1}
\\
\hspace{-0.6cm} \delta^{(1)} v^{\theta} &=& \alpha \Omega C_{l} l
    \left( \frac{r}{R} \right)^{l-1}
    \sin^{l-1}\theta \sin (l\phi+\omega t),
    \label{vf2}
\\
\hspace{-0.6cm} \delta^{(1)} v^{\phi} &=& \alpha \Omega  C_{l} l
    \left( \frac{r}{R} \right)^{l-1}
    \sin^{l-2}\theta \cos\theta
\cos (l\phi+\omega t), \label{vf3}
\end{eqnarray}
\end{subequations}
where $\delta^{(1)} v^i$ are the contravariant components of the
first-order Eulerian change in velocity, $R$ and $\Omega$ are,
respectively, the radius and the angular velocity of the
unperturbed star,
\begin{equation}
\omega=-\Omega l+ \frac{2\Omega}{l+1}
\end{equation}
is the mode's angular frequency in an inertial frame, $\alpha$ is
the mode's amplitude and $C_l$ is chosen to be
\begin{equation}
C_l=(2l-1)!!\sqrt{\frac{2l+1}{2\pi(2l)!l(l+1)}}.
\end{equation}
To the velocity field (\ref{vf}) corresponds a Lagrangian vector
displacement $\xi^{(1)i}$,
\begin{subequations}
\label{ld}
\begin{eqnarray}
\xi^{(1)r}&=&0,  \label{ld1}
\\
\xi^{(1)\theta}&=& - \frac12 \alpha C_l l(l+1) \left( \frac{r}{R}
\right)^{l-1} \sin^{l-1} \theta \cos (l\phi+\omega t), \label{ld2}
\nonumber
\\
& &
\\
\xi^{(1)\phi}&=& \frac12 \alpha C_l l(l+1) \left( \frac{r}{R}
\right)^{l-1} \sin^{l-2}\theta \cos\theta \sin(l\phi+\omega t),
\label{ld3} \nonumber
\\
& &
\end{eqnarray}
\end{subequations}
satisfying, at lowest order in $\Omega$, the equations
\begin{equation}
\delta^{(1)} v^i=\partial_{t} \xi^{(1)i} + v^k\nabla_k\xi^{(1)i} -
\xi^{(1)k}\nabla_k v^i
\end{equation}
and \begin{equation} \nabla_i(\rho \xi^{(1)i}) = 0,
\end{equation}
where $v^i=\Omega \delta_{\phi}^i$ and $\rho$ are, respectively,
the fluid velocity and the mass density of the unperturbed star.
The knowledge of the first-order vector displacement $\xi^{(1)i}$
is sufficient to compute the canonical angular momentum of the
\textit{r}-mode perturbation \cite{fs}; using Eq.~(\ref{ld}), one
obtains:
\begin{eqnarray}
J_c &=& - \int \rho \partial_{\phi} \xi^{(1)i} \left(
\partial_{t} \xi^{(1)}_i
+v^k \nabla_k \xi^{(1)}_i \right) dV \nonumber \\
&=& -\frac14 \alpha^2 \Omega l(l+1)  R^{2-2l} \int_0^R \rho
r^{2l+2} dr. \label{canangm}
\end{eqnarray}
However, the canonical angular momentum $J_c$ is not the full
physical angular momentum at second order $\delta^{(2)}J$, the
difference being given by \cite{fs}
\begin{equation}
\delta^{(2)}J- J_c=\frac{1}{\Omega}\int\rho v^i
\Delta_{\xi}^{(2)}v_i dV, \label{angm}
\end{equation}
where $\Delta_{\xi}^{(2)}v_i$ is the second-order Lagrangian
change in velocity
\begin{eqnarray}
\Delta^{(2)}_{\xi}v_i &=& \partial_{t} \xi^{(1)k} \nabla_i
\xi^{(1)}_k + v^k \nabla_k \xi^{(1)m}\nabla_i \xi^{(1)}_m
\nonumber
    \\
& & +\; \partial_{t} \xi^{(2)}_i + v^k
    \left( \nabla_i \xi^{(2)}_k + \nabla_k \xi^{(2)}_i \right),
\label{lc2v}
\end{eqnarray}
which, as opposed to the canonical angular momentum, also contains
terms linear in the second-order Lagrangian displacement vector
$\xi^{(2)a}$, in addition to terms quadratic in $\xi^{(1)a}$.
Therefore, the computation of the physical angular momentum at
second order requires the knowledge of $\xi^{(2)a}$.

In a recent paper \cite{sa}, a nonlinear extension of the linear
\textit{r}-mode was found within the nonlinear theory up to second
order in the mode's amplitude in the case of a slowly rotating,
Newtonian, barotropic, perfect-fluid star:
\begin{subequations}
\label{difrot}
\begin{eqnarray}
\delta^{(2)} v^r&=& 0,  \label{difrot1}
\\
\delta^{(2)} v^{\theta}&=&0,  \label{difrot2}
\\
\delta^{(2)} v^{\phi} &=&
  \frac12 \alpha^2 \Omega C_l^2 l^2(l^2-1)
  \left(\frac{r}{R} \right)^{2l-2} \sin^{2l-4}\theta \nonumber
\\
& & +\;  \alpha^2 \Omega A r^{N-1} \sin^{N-1}\theta,
  \label{difrot3}
\end{eqnarray}
\end{subequations}
where $A$ and $N$ are arbitrary constants fixed by initial data.
This second-order solution represents differential rotation,
producing large scale drifts of fluid elements along stellar
latitudes. The first term on the right-hand side of
Eq.~(\ref{difrot3}) is induced by first-order quantities, while
the second term is determined by the choice of initial data. Since
these two terms cannot cancel each other, differential rotation is
an unavoidable feature of \textit{r}-modes. Let us in the present
work restrict ourselves to the case $N=2l-1$. We also redefine $A$
to be
\begin{equation}
A=\frac12 C_l^2 l^2 (l+1) R^{2-2l} K, \label{parK}
\end{equation}
where $K$ is a constant fixed by the choice of initial data.

At lowest order in $\Omega$, the contravariant components of the
second-order Lagrangian displacement vector $\xi^{(2)i}$,
corresponding to the axisymmetric time-independent velocity field
(\ref{difrot}), are determined by the equations \cite{sa}
\begin{eqnarray}
\delta^{(2)} v^i  &=& \partial_{t} \xi^{(2)i} +
v^k\nabla_k\xi^{(2)i} - \xi^{(2)k}\nabla_k v^i \nonumber
\\
& & -\; \xi^{(1)k} \nabla_k\delta^{(1)} v^i, \label{vi2a}
\\
 \nabla_k \xi^{(2)k} &=& \frac12 \nabla_k \xi^{(1)m} \nabla_m
 \xi^{(1)k},
 \label{vi2b}
\\
\xi^{(2)k}\nabla_k\rho &=&
-\frac12\xi^{(1)k}\xi^{(1)m}\nabla_k\nabla_m\rho, \label{vi2c}
\end{eqnarray}
which yield the solution \cite{sa}:
\begin{subequations}
\label{edc}
\begin{eqnarray}
\xi^{(2)r} &=& \frac{1}{16} \alpha^2 C_l^2 l^2 (l+1)^2 R \left(
\frac{r}{R} \right)^{2l-1} \nonumber
\\
 & & \times \sin^{2l-2}\theta \left( \sin^2\theta-2 \right), \label{edc1}
\\
\xi^{(2)\theta} &=& \frac{1}{16} \alpha^2 C_l^2 l^2 (l+1)^2 \left(
\frac{r}{R} \right)^{2l-2} \nonumber
\\
& & \times \sin^{2l-3}\theta \cos\theta \left( \sin^2\theta+2l-2
\right), \label{edc2}
\\
\xi^{(2)\phi} &=& \frac14 \alpha^2 \Omega C_l^2 l^2 (l+1)(2K+2l-1)
\left( \frac{r}{R} \right)^{2l-2} \nonumber
\\
& & \times \sin^{2l-2}\theta \; t + C(r,\theta), \label{edc3}
\end{eqnarray}
\end{subequations}
where $C$ is an arbitrary function of $r$ and $\theta$. In
Eq.~(\ref{vi2a})--(\ref{vi2c}), terms quadratic in first-order
quantities give rise to double ``frequency" terms of the type
$\cos[2(l\phi+\omega t)]$ inducing a second-order solution
corresponding to an oscillating response at a ``frequency" twice
that of the linear \textit{r}-mode; however, in this article these
double frequency oscillations will not be considered since they
give a zero contribution to the second-order physical angular
momentum of the \textit{r}-mode perturbation. Using the above
expressions for the second-order Lagrangian displacement vector
$\xi^{(2)i}$, the physical angular momentum at second order is
computed to be \cite{sa}:
\begin{eqnarray}
\delta^{(2)} J &=& \frac12 \alpha^2 \Omega \left[ 2Kl+(l-1)(2l+1)
\right] R^{2-2l} \nonumber
\\
& & \times \int_0^R \rho r^{2l+2} dr. \label{physangm}
\end{eqnarray}

As can be seen from Eqs.~(\ref{canangm}) and (\ref{physangm}), the
physical angular momentum of the \textit{r}-mode perturbation is
not equal, in general, to the canonical angular momentum. The
former contains also a part linear in the second-order Lagrangian
change in velocity, which, as pointed out in Ref.~\cite{fs}, is
related to conservation of circulation in the fluid. Since, as
shown in Ref.~\cite{sa}, at second order \textit{r}-modes do not
conserve vorticity, it follows that, in general, the physical and
canonical angular momentum do not coincide. However, specific
choices of $K$ can be made such that the integral in
Eq.~(\ref{angm}) vanishes. Such a case ($K=-2$, for $l=2$) was
studied in Ref.~\cite{olcsva} within a phenomenological model for
the evolution of the \textit{r}-mode instability. There is not,
however, to our knowledge, a physical condition that forces $K$ to
take such a particular value. Therefore, in this paper we study
the evolution of the \textit{r}-mode instability for arbitrary
values of $K$.

\section{\label{model}The evolution model}

In this paper, the simple model proposed in Ref.~\cite{olcsva} to
study the evolution of the \textit{r}-mode perturbation in newly
born, hot, rapidly rotating neutron stars is adopted with two
modifications. First, the physical angular momentum of the
\textit{r}-mode perturbation is not taken to be just the canonical
angular momentum as in Ref.~\cite{olcsva}. Instead, we will use
the full physical angular momentum (\ref{physangm}), which, as
already discussed in the previous section, because of the presence
of differential rotation induced by \textit{r}-modes, is
different, in general, from the canonical angular momentum. A
second modification, less significant, is related to the proposal
of Ref.~\cite{hl} to deduce the evolution equations for the mode's
amplitude $\alpha$ and the star's angular velocity $\Omega$ just
from angular momentum considerations.

Since the most unstable \textit{r}-mode is the $l=2$ mode, we will
focus in this paper our attention only on the evolution of this
mode. We will also assume, as in Ref.~\cite{olcsva}, that the mass
density $\rho$ and the pressure $p$ of the fluid are related by a
polytropic equation of state $p=k\rho^2$, with $k$ such that
$M=1.4 M_{\bigodot}$ and $R=12.53$ km.

In our model, it is assumed that the total angular momentum of the
star is given by
\begin{equation}
J=I\Omega+\delta^{(2)}J
 \label{amtotal}
\end{equation}
where the momentum of inertia of the equilibrium configuration $I$
is given by
\begin{equation}
I = \frac{8\pi}{3} \int_0^R\rho r^4 dr  = \tilde{I}MR^2,
\end{equation}
with $\tilde{I}=0.261$, and the physical angular momentum of the
\textit{r}-mode perturbation $\delta^{(2)}J$ is given by
\begin{eqnarray}
\delta^{(2)} J &=& \frac12 \alpha^2 \Omega (4K+5) \frac{1}{R^2}
\int_0^R\rho r^6 dr \nonumber
\\
&=& \frac12 \alpha^2 \Omega (4K+5) \tilde{J} M R^2,
 \label{physangm2}
\end{eqnarray}
with $\tilde{J}=1.635\times10^{-2}$.

The total angular momentum of the star $J=J(\alpha,\Omega)$
decreases due to the emission of gravitational radiation according
to \cite{olcsva}
\begin{equation}
\frac{dJ}{dt}=3 \tilde{J} MR^2 \frac{\alpha^2\Omega}{\tau_{GR}},
\label{dim-mat}
\end{equation}
where the gravitational-radiation timescale $\tau_{GR}$ is given
by \cite{lom}
\begin{equation}
\frac{1}{\tau_{GR}} = \frac{1}{\tilde{\tau}_{GR}} \left(
\frac{\Omega^2}{\pi G \bar{\rho}} \right)^3, \label{gr-timescale}
\end{equation}
with the fiducial timescale $\tilde{\tau}_{GR}=-3.26\,\mbox{s}$.
In the above equation, $\bar{\rho}$ is the average mass density of
the star and $G$ is the gravitational constant.

From Eqs.~(\ref{amtotal})--(\ref{gr-timescale}) it is
straightforward to obtain a differential equation for the time
evolution of $\alpha$ and $\Omega$:
\begin{eqnarray}
& &\left[ 1+\frac{1}{3}(4K+5)Q\alpha^2 \right] \frac{d\Omega}{dt}
\nonumber
 \\
& & \hspace{1.0cm} +\; \frac{2}{3}(4K+5)Q\Omega\alpha
\frac{d\alpha}{dt}= 2Q\frac{\Omega\alpha^2}{\tau_{GR}},
\label{first-e}
\end{eqnarray}
where $Q=3\tilde{J}/(2\tilde{I})=0.094$.

Following the proposal of Ref.~\cite{hl} to deduce the evolution
equations just from angular momentum considerations, we assume
that the physical angular momentum of the \textit{r}-mode
perturbation $\delta^{(2)}J$ increases due to the emission of
gravitational radiation and decreases due to the dissipative
effect of viscosity,
\begin{equation}
\frac{d\delta^{(2)}J}{dt}=-2\delta^{(2)}J \left(
\frac{1}{\tau_{GR}} + \frac{1}{\tau_V} \right). \label{varpam}
\end{equation}
For the viscous timescale $\tau_V$, we take the expression derived
in Ref.~\cite{lom} for the simple case of the linear
\textit{r}-mode (\ref{vf}) of a newly born, hot, rapidly rotating
neutron star with shear and bulk viscosity:
\begin{equation}
\frac{1}{\tau_V} = \frac{1}{\tilde{\tau}_S} \left(
\frac{10^9\mbox{K}}{T} \right)^2 + \frac{1}{\tilde{\tau}_B} \left(
\frac{T}{10^9\mbox{K}} \right)^6 \left( \frac{\Omega^2}{\pi G
\bar{\rho}} \right), \label{v-timescale}
\end{equation}
with the fiducial timescales $\tilde{\tau}_S =
2.52\times10^8\,\mbox{s}$ and
$\tilde{\tau}_B=6.99\times10^8\,\mbox{s}$. Several authors have
taken into account other dissipative mechanisms, but in this paper
we shall restrict ourselves, for illustrative purposes, to the
above expression.

{} From Eq.~(\ref{varpam}) one obtains, using
Eq.~(\ref{physangm2}), a second differential equation for the time
evolution of $\alpha$ and $\Omega$:
\begin{equation}
2\Omega\frac{d\alpha}{dt}+\alpha\frac{d\Omega}{dt}=-2\alpha\Omega
\left( \frac{1}{\tau_{GR}} + \frac{1}{\tau_V} \right).
\label{second-e}
\end{equation}
Note that in Ref.~\cite{olcsva} it was assumed that it was the
energy of the perturbation (in the rotating frame), and not the
angular momentum, that increases due to emission of gravitational
waves and decreases due to viscosity. For a constant $\Omega$
these two approaches are coincident; however, for a varying
$\Omega$ they differ in quantities of the order of $Q\alpha^2$
\cite{hl}.

From Eqs.~(\ref{first-e}) and (\ref{second-e}) it is
straightforward to obtain a system of two, first-order, coupled,
differential equations determining the time evolution of the
amplitude of the \textit{r}-mode $\alpha(t)$ and of the angular
velocity of the star $\Omega(t)$:
\begin{eqnarray}
\frac{d\Omega}{dt} &=& \frac83 (K+2)Q
\frac{\Omega\alpha^2}{\tau_{GR}} + \frac23 (4K+5)Q
\frac{\Omega\alpha^2}{\tau_{V}}, \label{eqmod1}
\\
\frac{d\alpha}{dt} &=& -\left[ 1 + \frac43 (K+2)Q \alpha^2 \right]
\frac{\alpha}{\tau_{GR}} \nonumber \\
& & - \left[ 1 + \frac13 (4K+5)Q \alpha^2 \right]
\frac{\alpha}{\tau_{V}}. \label{eqmod2}
\end{eqnarray}

For $K=-2$, the above equations coincide with Eqs.~(2.7) and (2.8)
of Ref.~\cite{hl} (for $1/\tau_{M}=0$) and with Eqs.~(3.14) and
(3.15) of Ref.~\cite{olcsva} (up to terms of order $Q\alpha^2$).
In this paper, as already mentioned above, we study the evolution
of the \textit{r}-mode instability for an arbitrary value of $K$.

\section{\label{role}The role of differential rotation in the evolution
of \textit{r}-modes}

The condition $\tau_{GR}^{-1}(\Omega)+ \tau_V^{-1}(\Omega,T) = 0$
gives the stability curve, i.e., the set of points in a diagram
$(\Omega,T)$ for which the damping effect of viscosity balances
exactly the driving effect of gravitational radiation. For a newly
born, hot, rapidly rotating neutron star, there is an interval of
temperatures and angular velocities of the star for which the
gravitational timescale $\tau_{GR}$ is much smaller than the
viscous timescale $\tau_{V}$, implying that the evolution of
$\alpha$ and $\Omega$ is determined, in a good approximation, by
the equations:
\begin{eqnarray}
\frac{d\Omega}{dt} &=& \frac83 (K+2)Q
\frac{\Omega\alpha^2}{\tau_{GR}}, \label{omega}
\\
\frac{d\alpha}{dt} &=& -\left[ 1 + \frac43 (K+2)Q \alpha^2 \right]
\frac{\alpha}{\tau_{GR}}. \label{alfa}
\end{eqnarray}
These equations remain a good approximation to Eqs.~(\ref{eqmod1})
and (\ref{eqmod2}) for a period of time during which the
temperature and angular velocity of the neutron star are such that
the corresponding point, in a $(\Omega,T)$ diagram, lies well
above the stability curve. For the model we have been considering,
this period of time is about one year \cite{olcsva}.

In solving the above system of equations, initial conditions at
$t=t_0$ are chosen such that $|\delta^{(2)}J(t_0)| \ll
I\Omega(t_0)$. For $\alpha_0\equiv\alpha(t_0)=10^{-6}$, this
implies that $|K|\ll 10^{13}$. In the case $K<-5/4$, for which
$\delta^{(2)}J<0$, as the amplitude of the mode grows due to the
gravitational-radiation instability, the total angular momentum of
the star, given by Eq.~(\ref{amtotal}), decreases and eventually
becomes negative. To avoid this, the growth of the mode's
amplitude has to be stopped by hand at a saturation value
$\alpha_{sat}\leqslant\sqrt{-3/[(4K+5)Q]}$; integration is then
carried on with a new set of equations for which
$\alpha=\alpha_{sat}$ and the evolution of the angular velocity
$\Omega$ is determined from Eq.~(\ref{first-e}), with
$d\alpha/dt=0$,
\begin{equation}
\frac{d\Omega}{dt}=\frac{2\Omega}{\tau_{GR}}
\frac{\alpha_{sat}^2Q}{1+\frac13(4K+5)\alpha_{sat}^2Q}.
\label{first-e1}
\end{equation}
In the case $K\geqslant-5/4$, for which $\delta^{(2)}J\geqslant0$,
as the amplitude of the mode grows, the total angular momentum
decreases but remains always positive and, therefore, there is no
need to saturate the mode by hand. In this case, $\alpha(t)$ and
$\Omega(t)$ are determined solely by Eqs.~(\ref{omega}) and
(\ref{alfa}) and, as will be shown below, the mode saturates
naturally at a value that depends on the parameter $K$, namely,
$\alpha_{sat}\propto(K+2)^{-1/2}$. As already mentioned above, the
parameter $K$, introduced in Eq.~(\ref{parK}), is fixed by initial
data, giving the initial amount of differential rotation
associated to the \textit{r}-mode. Thus, depending on the choice
of initial data, the saturation amplitude of the \textit{r}-mode
can be of order unity, as assumed in the first models of evolution
of the \textit{r}-mode instability \cite{olcsva}, or very small,
as recent studies within the nonlinear theory seem to imply
\cite{afmstw}.

The requirement that the total angular momentum is always
non-negative, implies that the right-hand side of Eq.~(\ref{alfa})
is always positive, i.e., $\alpha(t)$ increases in time for any
value of $K$. It is also worth mentioning that, according to
Eq.~(\ref{omega}), for $K\neq-2$, $\Omega(t)$ evolves on a
gravitational timescale, that is, a change in the background
motion of the star occurs simultaneously to the growth of the mode
due to the gravitational-wave instability. This result is not
completely unexpected, since such kind of behavior has already
been reported in the case of a toy model of a thin spherical shell
of a rotating incompressible fluid \cite{lu}.

Let us first consider the case $K\geqslant-5/4$. As already
mentioned above, in this case the total angular momentum of the
star is always positive and the amplitude of the mode $\alpha(t)$
and the angular velocity of the star $\Omega(t)$ are determined
solely by the system of Eqs.~(\ref{omega}) and (\ref{alfa}), from
which it is straightforward to obtain:
\begin{equation}
\frac{d\Omega}{\Omega} =
-\frac{\frac83(K+2)Q\alpha}{1+\frac43(K+2)Q\alpha^2} d\alpha.
\end{equation}
This differential equation yields the solution
\begin{equation}
\frac{\Omega}{\Omega_0} =
\frac{1+\frac43(K+2)Q\alpha_0^2}{1+\frac43(K+2)Q\alpha^2},
\label{omega-alfa}
\end{equation}
where $\Omega_0\equiv\Omega(t_0)$ and $\alpha_0\equiv\alpha(t_0)$
are, respectively, the initial angular velocity of the star and
the initial amplitude of the \textit{r}-mode perturbation. Using
the above expression to eliminate $\Omega$ from Eq.~(\ref{alfa})
and integrating, one obtains:
\begin{eqnarray}
& &-\frac{1}{\tilde{\tau}_{GR}}\left( \frac{\Omega_0}{\sqrt{\pi
G\bar{\rho}}} \right)^6 \left[ 1+\frac43(K+2)Q\alpha_0^2 \right]^6
(t-t_0) \nonumber
\\
& & \quad \quad \quad =\ln \frac{\alpha}{\alpha_0} + \sum_{n=1}^5
\frac{5!}{2n(5-n)!n!}\left[\frac43(K+2)Q\right]^n \nonumber
\\
& & \quad \quad \quad \quad \times \left( \alpha^{2n}-
\alpha_0^{2n} \right). \label{alfa-t}
\end{eqnarray}

In the initial stages of the evolution of the \textit{r}-mode
instability, the right-hand side of Eq.~(\ref{alfa-t}) is
dominated by the first term and $\alpha$ increases exponentially,
\begin{equation}
\alpha(t) \simeq \alpha_0 \exp \left\{ 0.027 \left(
\frac{\Omega_0}{\Omega_K} \right)^6 (t-t_0) \right\},
\label{l-fase}
\end{equation}
as expected; for later times, the right-hand side of
Eq.~(\ref{alfa-t}) is dominated by the last term ($n=5$) and
$\alpha$ increases very slowly as
\begin{equation}
\alpha(t) \simeq 2.48 \left( \frac{\Omega_0}{\Omega_K}
\right)^{3/5} (t-t_0)^{1/10}\frac{1}{\sqrt{K+2}}. \label{m-fase}
\end{equation}
In  Eqs.~(\ref{l-fase}) and (\ref{m-fase}), $t-t_0$ is given in
seconds, $\Omega_K=(2/3)\sqrt{\pi G\bar{\rho}}$ is the Keplerian
angular velocity at which the star starts shedding mass at the
equator, we have used for the gravitational timescale the value
$\tilde{\tau}_{GR}=-3.26\mbox{ s}$, and we have assumed that $K$
and $\alpha_0$ are such that $4(K+2)Q\alpha_0^2/3 \ll 1$. The
smooth transition between the regimes (\ref{l-fase}) and
(\ref{m-fase}) occurs for $t-t_0\simeq\mbox{few}\times 10^2$
seconds.

\begin{figure}[h]
\includegraphics{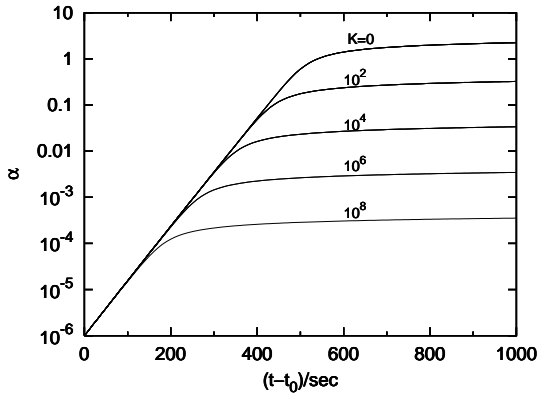}
\caption{\label{fig:alfa}Time evolution of the amplitude of the
\textit{r}-mode $\alpha$ for different values of $K$. The
saturation value of the amplitude of the mode depends crucially on
the parameter $K$, which is related to the amount of initial
differential rotation associated with the \textit{r}-mode. The
initial values of the amplitude of the mode and of the angular
velocity of the star are, respectively, $\alpha_0=10^{-6}$ and
$\Omega_0=\Omega_K$.}
\end{figure}

The equation~(\ref{alfa-t}) describes the time evolution of the
amplitude of the mode between the moment the gravitational-wave
instability sets in and the moment viscosity effects become
dominant and start damping the mode. It reveals that saturation of
the \textit{r}-mode occurs just a few hundred seconds after the
beginning of the exponential growth of $\alpha$. Thus, within the
model of evolution we are studying, which includes the nonlinear
effect of the differential rotation induced by \textit{r}-modes,
the mode's amplitude saturates in a natural way (see
Fig.~\ref{fig:alfa}).

The saturation value of the amplitude of the mode depends
crucially on the parameter $K$, $\alpha_{sat}\propto
(K+2)^{-1/2}$. The initial amount of differential rotation
associated with the \textit{r}-mode can be minimized by choosing
$K$ appropriately, namely, $K\simeq0$, with the consequence that
the amplitude of the mode saturates at values of order unity. In
this case, other nonlinear effects as, for instance, mode--mode
interactions, that saturate \textit{r}-modes at values much
smaller than unity \cite{afmstw}, are more relevant than
differential rotation in what concerns the saturation amplitude of
the mode. However, if the initial differential rotation associated
to \textit{r}-modes is significant, when compared with
$\alpha_0^{-1}$, then the saturation amplitude $\alpha_{sat}$ can
be as small as $10^{-3}-10^{-4}$. In this case, differential
rotation plays an important role in the saturation of
\textit{r}-modes.

Let us now turn our attention to the time evolution of the angular
velocity of the star $\Omega$. Integrating Eq.~(\ref{omega}),
after eliminating $\alpha$ with Eq.~(\ref{omega-alfa}), one
obtains the following solution:
\begin{eqnarray}
& & -\frac{2}{\tilde{\tau}_{GR}}\left( \frac{\Omega_0}{\sqrt{\pi
G\bar{\rho}}}
\right)^6 \left[ 1+\frac43(K+2)Q\alpha_0^2 \right]^6 (t-t_0) \nonumber \\
& & \quad \quad \quad = \sum_{n=1}^5
\frac{\left[1+\frac43(K+2)Q\alpha_0^2\right]^n}{n}
\left[ \left( \frac{\Omega_0}{\Omega} \right)^n-1 \right] \nonumber \\
& & \quad \quad \quad \quad +\; \ln \frac{\Omega_0}{\Omega} + \ln
\frac{1+\frac43(K+2)Q\alpha_0^2 -
\frac{\Omega}{\Omega_0}}{\frac43(K+2)Q\alpha_0^2}. \label{omega-t}
\end{eqnarray}

\begin{figure}[h]
\includegraphics{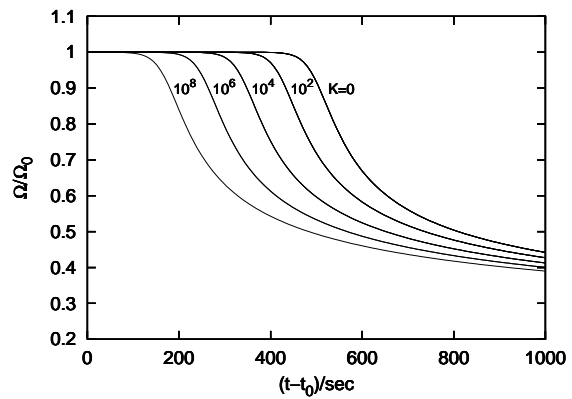}
\caption{Time evolution of the angular velocity of the star
$\Omega$ for different values of $K$. After a few hundred seconds,
the value of the angular velocity becomes quite insensitive to the
value of $K$. The initial values of the amplitude of the mode and
of the angular velocity of the star are, respectively,
$\alpha_0=10^{-6}$ and $\Omega_0=\Omega_K$.} \label{fig:omega}
\end{figure}

In the initial stages of the evolution of the \textit{r}-mode
instability, the right-hand side of Eq.~(\ref{omega-t}) is
dominated by the last term and $\Omega$ decreases as
\begin{equation}
\frac{\Omega(t)}{\Omega_0} \simeq 1-\frac43(K+2)Q\alpha_0^2 \exp
\left\{ 0.054 \left( \frac{\Omega_0}{\Omega_K} \right)^6 (t-t_0)
\right\}; \label{i-fase}
\end{equation}
for later times, the right-hand side of Eq.~(\ref{omega-t}) is
dominated by the first term ($n=5$) and $\Omega$ decreases slowly
as
\begin{equation}
\frac{\Omega(t)}{\Omega_0} \simeq  1.30 \left(
\frac{\Omega_0}{\Omega_K} \right)^{-6/5} (t-t_0)^{-1/5}.
\label{f-fase}
\end{equation}
In Eqs.~(\ref{i-fase}) and (\ref{f-fase}), again, $t-t_0$ is given
in seconds, we have used for the gravitational timescale the value
$\tilde{\tau}_{GR}=-3.26\mbox{ s}$, and we have assumed that $K$
and $\alpha_0$ are such that $4(K+2)Q\alpha_0^2/3 \ll 1$. The
smooth transition between the regimes (\ref{i-fase}) and
(\ref{f-fase}) occurs for $t-t_0\simeq\mbox{few}\times 10^2$
seconds (see Fig.~\ref{fig:omega}).
Remarkably, in the later phase of the evolution, the angular
velocity $\Omega$ does not depend on the value of $K$, and,
consequently, does not depend on the saturation value of $\alpha$.
As can be seen in Fig.~\ref{fig:omega}, already at
$t-t_0=1000\,\mbox{s}$, for values of $K$ ranging from 0 to
$10^8$, the angular velocities are not much different; for
$t-t_0\simeq1\,\mbox{year}$ any difference becomes negligibly
small. This contrasts with the results obtained in
Ref.~\cite{olcsva}, where the value of $\Omega$ after about one
year depends critically on the choice of $\alpha_{sat}$ (see case
$K=-2$ below).

After about one year of evolution, when the dissipative effect of
viscosity becomes dominant and starts damping the mode, the
angular velocity of the star reaches values consistent with
observational results. Indeed, assuming that initially the star
rotates with the maximum allowed angular velocity,
$\Omega_0=\Omega_K=(2/3)\sqrt{\pi G \bar{\rho}}$, one obtains from
Eq.~(\ref{omega-t}) that $\Omega_{\mbox{\small{one year}}}\simeq
0.042 \Omega_K$, in good agreement with the inferred initial
angular velocity of the fastest pulsars associated with supernova
remnants.

\begin{figure}[h]
\includegraphics{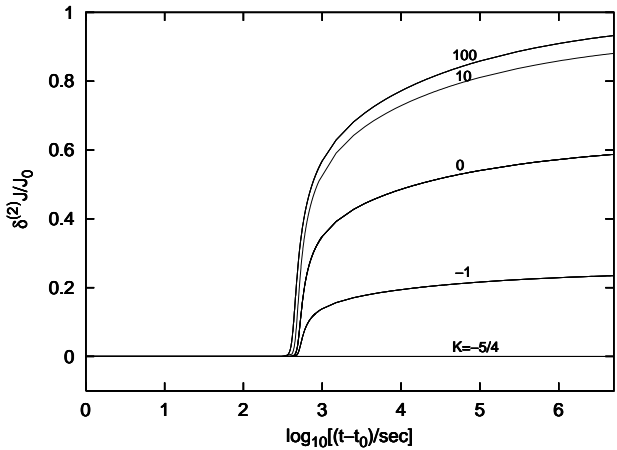}
\caption{Time evolution of the physical angular momentum of the
\textit{r}-mode perturbation $\delta^{(2)}J$ for different values
of $K$. For $K\gg1$, most of the initial angular momentum of the
star $J_0\simeq I\Omega_0$ is transferred to the \textit{r}-mode
perturbation and, consequently, almost none is carried away by
gravitational radiation emission. The initial values of the
amplitude of the mode and of the angular velocity of the star are,
respectively, $\alpha_0=10^{-6}$ and $\Omega_0=\Omega_K$.}
\label{fig:jota2l}
\end{figure}

\begin{figure}[h]
\includegraphics{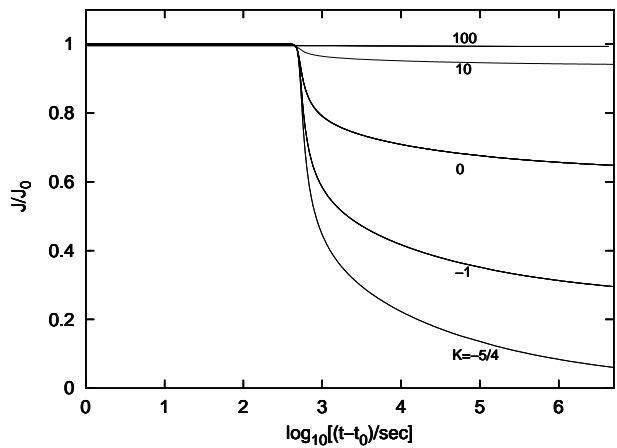}
\caption{Time evolution of the total angular momentum of the star
$J$ for different values of $K$. For small values of $K$, a
significant part of the initial angular momentum of the star is
radiated away by gravitational radiation. The initial values of
the amplitude of the mode and of the angular velocity of the star
are, respectively, $\alpha_0=10^{-6}$ and $\Omega_0=\Omega_K$.}
\label{fig:jota}
\end{figure}

From Eqs.~(\ref{m-fase}) and (\ref{f-fase}) follows that, for
$t\gg \mbox{few}\times10^2$ seconds, the quantity
$(\Omega/\Omega_0)\alpha^2$ does not depend on time, just on $K$,
namely, $(\Omega/\Omega_0)\alpha^2\simeq8/(K+2)$, implying that
the fraction of the initial angular momentum of the star that is
transferred to the \textit{r}-mode is just a function of $K$,
\begin{equation}
\delta^{(2)}J/J_0\simeq  \frac{K+5/4}{K+2}, \quad \mbox{for } t\gg
\mbox{few}\times10^2 \mbox{ seconds},
\end{equation}
where we have taken $J_0\simeq I\Omega_0$. Thus, for $K\gg1$, most
of the initial angular momentum of the star $I\Omega_0$ is
transferred to the \textit{r}-mode perturbation and, consequently,
almost none is carried away by gravitational radiation emission.
On the other hand, for $K=-5/4$, the angular momentum of the
perturbation $\delta^{(2)}J$ remains zero during the evolution
and, consequently, all the initial angular momentum of the star
$I\Omega_0$ is available to be radiated away by gravitational
radiation (see Figs.~\ref{fig:jota2l} and \ref{fig:jota}).

\begin{figure}[h]
\includegraphics{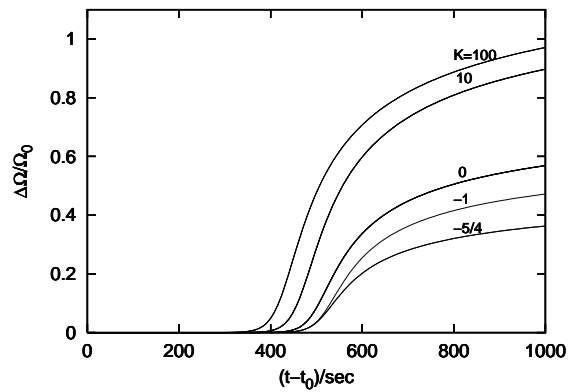}
\caption{Time evolution of the average differential rotation
$\Delta\Omega$ for different values of $K$. After a few hundred
seconds the average differential rotation increases rapidly,
saturating at high values relatively to the initial angular
velocity of the star. The initial value of the amplitude of the
mode is $\alpha_0=10^{-6}$.} \label{fig:Delta-omega}
\end{figure}

During the nonlinear evolution, the fluid develops a strong
differential rotation. Let us define the average differential
rotation $\Delta\Omega$ as the weighted variance of
$\Omega$~\cite{ltv2},
\begin{equation}
(\Delta\Omega)^2=\frac{\int\rho r^2 \sin^2\theta
(\delta^{(2)}v^\phi-\bar{\Omega}_{dr})^2 dV}{\int\rho r^2
\sin^2\theta dV}, \label{aver-rot-dif}
\end{equation}
where the average angular velocity $\bar{\Omega}_{dr}$,
characterizing the drift of fluid elements along stellar
latitudes, is given by
\begin{equation}
\bar{\Omega}_{dr} =\frac{\delta^{(2)}J}{I} = \frac{\int\rho r^2
\sin^2\theta \delta^{(2)}v^\phi dV}{\int\rho r^2 \sin^2\theta dV}.
\label{aver-ang-vel}
\end{equation}
For the polytropic model we have been considering,
Eqs.~(\ref{aver-rot-dif}) and (\ref{aver-ang-vel}) yield for the
average differential rotation the following expression:
\begin{eqnarray}
\Delta\Omega &=& \frac13 \alpha^2 \Omega Q {\biggl [}
\frac{15}{56} \left( 24K^2+56K+35 \right)
\frac{\tilde{I}\tilde{H}}{\tilde{J}^2} \nonumber
\\
& & -\; (4K+5)^2 {\biggr ]}^{1/2}, \label{aver-rot-dif-n}
\end{eqnarray}
where $\tilde{H}=\int_0^R \rho(r) r^8 dr/(M R^6)=0.01$. As can be
seen from Fig.~\ref{fig:Delta-omega}, after a few hundred seconds
the average differential rotation increases rapidly, saturating at
high values relatively to the initial angular velocity of the
star.
Note that the average differential rotation never vanishes for any
value of $K$. This is a consequence of the fact that the
second-order velocity field (\ref{difrot}) has two components, one
induced by first-order quantities and another, fixed by initial
data, which is a pure second-order effect. Since these two terms
cannot cancel each other, a velocity drift of fluid elements along
stellar latitudes is an unavoidable feature of the nonlinear
\textit{r}-mode pulsation.

Let us now turn to the case $K<-5/4$. As already mentioned above,
at a certain point of the evolution, the \textit{r}-mode has to be
saturated by hand in order to avoid that the total angular
momentum of the star becomes negative. Thus, during a first stage
of the evolution, $\alpha$ and $\Omega$ are determined from
Eqs.~(\ref{omega}) and (\ref{alfa}) and during a second stage,
$\alpha=\alpha_{sat}\leqslant\sqrt{-3/[(4K+5)Q]}$ and $\Omega$ is
determined by Eq.~(\ref{first-e1}). The fact that a saturation
value for $\alpha$ has to be fixed by hand introduces an element
of arbitrariness which did not exist in the case $K\geqslant-5/4$.
In particular, by choosing appropriately $\alpha_{sat}$ it is
possible to arrange that the final value of $\Omega$ is consistent
with observational data and, additionally, that a significant part
of the initial angular momentum of the star is carried away by
gravitational waves.

For $-2<K<-5/4$ and $K<-2$, the solution of the system of
equations (\ref{omega}) and (\ref{alfa}) is given by
Eqs.~(\ref{alfa-t}) and (\ref{omega-t}). During the second stage
of evolution, $\alpha=\alpha_{sat}\leqslant\sqrt{-3/[(4K+5)Q]}$
and $\Omega(t)$ is given by
\begin{eqnarray}
\Omega(t) &=& \Omega(t_*) {\biggl [}
\frac{0.030\alpha_{sat}^2}{1+\frac13(4K+5) Q \alpha_{sat}^2}
\nonumber
\\
& & \times \left( \frac{\Omega(t_*)}{\Omega_K} \right)^6 (t-t_*)
 + 1 {\biggr ]}^{-1/6}, \label{omega-k54}
\end{eqnarray}
where $t_*$ is the time at which occurs the transition from the
first to the second stage of evolution and $t-t_*$ is given in
seconds. During the two stages of evolution we have just
described, the total angular momentum of the star $J(t)$ decreases
due to gravitational-wave emission. At the end of the second stage
of the evolution, it is given by
\begin{equation}
J(t_{final}) \simeq J_0 \left[ 1 + \frac13 (4K+5)Q \alpha_{sat}^2
\right] \frac{\Omega(t_{final})}{\Omega_0}, \label{fam}
\end{equation}
where we have taken $J_0\simeq I\Omega_0$ and $\Omega(t)$ is given
by Eq.~(\ref{omega-k54}). Finally, let us point out that, for
$K<-2$, during the first stage of evolution, the angular velocity
of the star increases.

The case $K=-2$ has to be treated separately. It corresponds to
the model studied in detail in Ref.~\cite{olcsva}. The system of
equations (\ref{omega}) and (\ref{alfa})  yields the solution
$\Omega=\Omega_0$ and $\alpha=\alpha_0
\exp\{-(t-t_0)/\tau_{GR}\}$. If the initial angular velocity is
chosen to be $\Omega_0=\Omega_K$, then
$\tau_{GR}=-37.1\,\mbox{s}$, implying that the perturbation grows
exponentially from the initial amplitude $\alpha_0=10^{-6}$ to
values of the order unity in just about $500\,\mbox{s}$
\cite{olcsva}. After this short initial period in which $\Omega$
is constant and $\alpha$ grows exponentially, the amplitude
$\alpha$ has to be forced, by hand, to take a certain saturation
value $\alpha_{sat}\leqslant Q^{-1/2}=3.26$, and the angular
velocity of the star is then given by Eq.~(\ref{omega-k54}), with
$K=-2$ and $\Omega(t_*)=\Omega_0$. The final angular momentum of
the star is given by Eq.~(\ref{fam}) with $K=-2$. As can be seen
from these equations, the final angular velocity and momentum of
the star depend critically on the saturation value of the mode's
amplitude $\alpha_{sat}$; for instance, after one year of
evolution, for $\alpha_{sat}=1$ and $\Omega_0=\Omega_K$ one
obtains $\Omega\simeq 0.1\Omega_K$ and $J\simeq 0.09 J_0$, while
for $\alpha_{sat}=10^{-3}$ the angular velocity and momentum are,
respectively, $\Omega\simeq 0.9\Omega_K$ and $J\simeq 0.9 J_0$.

\section{\label{con}Conclusions}

In this paper we have studied the role of differential rotation in
the evolution of the \textit{r}-mode instability. We have adopted
the simple model of Ref.~\cite{olcsva}, with two modifications:
(i) the physical angular momentum of the \textit{r}-mode
perturbation includes not only the canonical angular momentum but
also a piece linear in second-order quantities, corresponding to
differential rotation inducing large scale drifts of fluid
elements along stellar latitudes; (ii) the evolution equations are
deduced just from angular momentum considerations. The first
modification is a quite important one, resulting from the fact
that differential rotation is an unavoidable kinematic feature of
\textit{r}-modes \cite{sa}. The presence of this differential
rotation implies that at second order in the mode's amplitude
\textit{r}-modes do not preserve vorticity of fluid elements. This
in turn implies that the canonical angular momentum is not the
full angular momentum at second order; one should also include a
part linear in the second-order Lagrangian change in velocity,
which is related to conservation of circulation in the fluid. The
second modification, less significant, leads to a system of
equations for the evolution of the \textit{r}-mode instability
that differs from the ones of Ref.~\cite{olcsva} just in a
quantity of the order of $Q\alpha^2$. A detailed justification for
this modification can be found in Ref.~\cite{hl}.

Within this model, we have derived a system of two first-order,
coupled, differential equations (\ref{eqmod1}) and (\ref{eqmod2}),
determining the time evolution of the amplitude of the
\textit{r}-mode $\alpha(t)$ and the angular velocity of the star
$\Omega(t)$. For the gravitational and viscous timescales
appearing in these equations, we have used the expressions derived
in Refs.~\cite{lom,aks} for the simple case of \textit{r}-modes of
a newly born, hot, rapidly rotating neutron star with shear and
bulk viscosity. In this case, the driving effect of the
gravitational radiation reaction overcomes the damping effect of
shear and bulk viscosity for about one year, while the temperature
of the star decreases from about $10^{10}$ K to about $10^9$ K.
During this period of time, in which the \textit{r}-mode
instability is active, the gravitational timescale is much smaller
than the viscous timescale, $\tau_{GR}\ll\tau_V$, and, therefore,
the evolution of $\alpha$ and $\Omega$ can be determined, in a
good approximation, by the system of equations (\ref{omega}) and
(\ref{alfa}).

The system of equations (\ref{omega}) and (\ref{alfa}) contains a
parameter $K$, which is fixed by initial data and gives the
initial amount of differential rotation associated with the
\textit{r}-mode. The specific case $K=-2$, for which the physical
angular momentum of the \textit{r}-mode perturbation coincides
with the canonical angular momentum, was studied in great detail
in Ref.~\cite{olcsva}. There is not, however, to our knowledge,
any physical condition that forces $K$ to take such a particular
value $K=-2$. Therefore, we have solved, both numerically and
analytically, the system of equations (\ref{omega}) and
(\ref{alfa}) for arbitrary $|K|\ll 10^{13}$. This upper limit for
$K$ results from the fact that one wishes to impose the condition
that the initial absolute value of the physical angular momentum
of the perturbation $|\delta^{(2)}J(t_0)|$ is much smaller than
the angular momentum of the unperturbed star $I\Omega_0$. In terms
of the initial drift of fluid elements along stellar latitudes,
this condition implies that $(v_{dr})_0\ll\Omega_0R$ at the
equator of the star. It could be argued that some dissipative
mechanism, such as bulk viscosity or magnetic coupling, would
reduce differential rotation in a newly born, hot, rapidly
rotating neutron star, before the \textit{r}-mode instability sets
in, implying that $K\simeq0$. However, even in that case, a
residual differential rotation of \textit{r}-modes would be
present in the star and the average differential rotation
$\Delta\Omega$ would increase exponentially as the amplitude of
the \textit{r}-mode grows, saturating at values relatively high as
compared with the initial angular velocity of the star. Finally,
in what concerns the admissible values of $K$, let us point out
that for $K<-5/4$, as the amplitude of the mode grows due to the
gravitational-radiation instability, the total angular momentum of
the star decreases and eventually becomes negative. To avoid this
unphysical situation, the growth of the mode's amplitude has to be
stopped by hand at a saturation value
$\alpha_{sat}\leqslant\sqrt{-3/[(4K+5)Q]}$ and integration has
then to be carried on with a new set of equations. The fact that a
saturation value for $\alpha$ has to be fixed by hand introduces
an element of arbitrariness into the solution, permitting, for
instance, that agreement between the predicted final value of the
angular velocity of the star and the value inferred from
astronomical observations can always be achieved by simply fine
tuning the value of $\alpha_{sat}$. One would wish, of course,
that such an agreement, even qualitative, would arise in a natural
way, without fine tuning the parameters of the model. For this
reason, we have concentrated most of our attention on the case
$K\geqslant-5/4$, in which such arbitrariness does not exist.

{} From the exact analytical solution (\ref{alfa-t}) and
(\ref{omega-t}) of the system of equations (\ref{omega}) and
(\ref{alfa}) one can extract several conclusions.

First, the amplitude of the \textit{r}-mode saturates in a natural
way a few hundred seconds after the mode instability sets in. The
saturation amplitude depends on the parameter $K$, namely,
$\alpha_{sat}\propto(K+2)^{-1/2}$. Therefore, if the initial
differential rotation of \textit{r}-modes is small ($K\simeq0$),
then the \textit{r}-mode saturates at values of the order of
unity. On the other hand, if the initial differential rotation is
significant ($K\gg1$), then the saturation amplitude can be as
small as $10^{-3}-10^{-4}$. These low values for the saturation
amplitude of \textit{r}-modes are of the same order of magnitude
as the ones obtained in recent investigations on wind-up of
magnetic fields \cite{rls} and on nonlinear mode-mode interaction
\cite{afmstw}.

Second, the value of the angular velocity of the star becomes,
after a short period of evolution ($t\gg \mbox{ few}\times 10^2$
s), very insensitive to the value of the parameter $K$, i.e., it
becomes insensitive to the saturation value of the mode's
amplitude. From Eq.~(\ref{omega-t}) for $t=t(\Omega)$, it can be
easily obtained that the angular velocity after about one year of
evolution of the \textit{r}-mode instability is $0.042\Omega_K$
(for any $K$), in good agreement with the inferred initial angular
velocity of the fastest pulsars associated with supernova
remnants.

Finally, the value of the physical angular momentum
$\delta^{(2)}J$ of the \textit{r}-mode perturbation tends, after a
short period of evolution ($t\gg \mbox{ few}\times 10^2$ s), to a
constant value given approximately by $[(K+\frac54)/(K+2)]J_0$.
Thus, for $K\gg1$ most of the initial angular momentum of the star
is transferred to the \textit{r}-mode and, consequently, almost
none is carried away by gravitational radiation. That the
\textit{r}-mode absorbs most of the initial angular momentum of
the star is explained by the fact that the mode develops a very
strong differential rotation. Indeed, the average differential
rotation, given by Eq.~(\ref{aver-rot-dif-n}), increases
exponentially, saturating after a few hundred seconds of evolution
at very high values relatively to the initial angular velocity of
the star. On the other hand, for values of $K\simeq-5/4$, the
transfer of angular momentum to the \textit{r}-mode is less strong
and, consequently, more angular momentum is available to be
carried away by gravitational radiation.

As we have seen, differential rotation introduces in the
evolutionary picture of the \textit{r}-mode instability new and
somewhat unexpected features. This differential rotation has a
kinematic origin, being induced by terms quadratic in the velocity
field of the linear \textit{r}-mode. However, differential
rotation of \textit{r}-modes could also be induced by the
gravitational-radiation reaction force. Indeed, within a toy model
of a thin spherical shell of a rotating incompressible fluid it
was shown that differential rotation of \textit{r}-modes is also
driven by a gravitational radiation force, leading to the
conjecture that in real stars one could observe a similar behavior
\cite{lu}. The existence of this differential rotation induced by
gravitational radiation and its influence on the evolution of the
\textit{r}-mode instability will be addressed in a future
publication \cite{ds}.

\begin{acknowledgments}
We thank \'Oscar Dias for helpful discussions. This work was
supported in part by the Funda\c c\~ao para a Ci\^encia e a
Tecnologia (FCT), Portugal. BT acknowledges financial support from
FCT through grant PRAXIS XXI/BD/21256/99.
\end{acknowledgments}

\bibliography{evolution}

\end{document}